\documentclass[10pt]{article}
\usepackage[dvips]{graphicx}
\usepackage{amsmath}
\usepackage{amsthm}
\usepackage{amssymb}
\usepackage{makeidx}

\newtheorem*{proposition}{Proposition}
\newtheorem*{corolary}{Corolary}
\newtheorem*{definition}{Definition}
\newtheorem*{example}{Example}

\title{Quantum Error Correction of Time-Correlated Errors}

\author{Feng Lu and Dan C. Marinescu\\
 School of Electrical Engineering and Computer Science \\
 University of Central Florida,  P. O. Box 162362\\
 Orlando, FL 32816-2362\\
 Email: { (lvfeng, dcm)@cs.ucf.edu}}

\begin{document}
\maketitle
\date
\begin{abstract}
The complexity of the error correction circuitry forces us to
design quantum error correction codes capable of correcting a
single error per error correction cycle. Yet, time-correlated
error are common for physical implementations of quantum systems;
an error corrected during the previous cycle may reoccur later due
to physical processes specific for each physical implementation of
the qubits. In this paper we study quantum error correction for a
restricted class of time-correlated errors in a spin-boson model.
The algorithm we propose allows the correction of two errors per
error correction cycle, provided that one of them is
time-correlated. The algorithm can be applied to any stabilizer
code when the two logical qubits $\mid 0_L \rangle$ and $\mid 1_L
\rangle$ are entangled states of $2^{n}$ basis states in
$\mathcal{H}_{2^n}$.

\end{abstract}

\section{Quantum Error Correction}
\label{QuantumErrorCorrection}

Quantum states are subject to decoherence and the question whether
a reliable quantum computer could be built was asked early on. A
``pure state'' $\mid \varphi \rangle = \alpha_{0} \mid 0 \rangle +
\alpha_{1} \mid 1 \rangle$ may be transformed as a result of the
interaction with the environment into a ``mixed state'' with
density matrix:

$$\rho = \mid \alpha_{0} \mid^2 \mid 0 \rangle \langle 0 \mid +
\mid \alpha_{1} \mid^2 \mid 1 \rangle \langle 1 \mid.
$$
Other forms of decoherence, e.g. leakage may affect the state
probability amplitude as well.

The initial thought was that a quantum computation could only be
carried out successfully if its duration is shorter than the
decoherence time of the quantum computer.  As we shall see in
Section \ref{TimeCorrelatedErrors}, the decoherence time ranges
from about $10^4$ seconds for the nuclear spin embodiment of a
qubit, to $10^{-9}$ seconds for quantum dots based upon charge.
Thus, it seemed very problematic that a quantum computer could be
built unless we have a mechanism to deal periodically with errors.
Now we know \cite{Shor96b} that quantum error correcting codes can
be used to ensure fault-tolerant quantum computing; quantum error
correction allows us to deal algorithmically with decoherence.
There is a significant price to pay to achieve fault-tolerance
through error correction: the number of qubits required to correct
errors could be several orders of magnitude larger than the number
of ``useful'' qubits \cite{DiVincenzo00}.

We describe the effect of the environment upon a qubit as a
transformation given by Pauli operators: (i) the state of the
qubit is unchanged if we apply the  $\sigma_{I}$ operator; (ii) a
bit-flip error is the result of applying the transformation given
by $\sigma_{x}$; (iii) a phase-flip error is the result of
applying the transformation given by $\sigma_{z}$; and (iv) a
bit-and-phase flip error is the result of applying the
transformation given by $\sigma_{y}= i \sigma_{x} \sigma_{z} $.

A quantum error correcting scheme takes advantage of entanglement
in two ways:

\begin{itemize}

\item We entangle one qubit carrying information with $(n-1)$
other qubits initially in state $\mid 0 \rangle$ and create an
$n$-qubit quantum codeword which is more resilient to errors.

\item We entangle the $n$ qubits of the quantum codeword with
ancilla qubits in such a way that we can measure the ancilla
qubits to determine the error syndrome without altering the state
of the $n$-qubit codeword, by performing a so called non
demolition measurement. The error syndrome tells if the individual
qubits of the codeword have been affected by errors as well as the
type of error.

\item
Finally, we correct the error(s).

\end{itemize}

Even though the no-cloning theorem prohibits the replication of a
quantum state, we are able to encode a single logical qubit as
multiple physical qubits and thus we can correct quantum errors
\cite{Steane96}. For example, we can encode the state of a qubit
$$
\mid \psi \rangle = \alpha_{0} \mid 0 \rangle + \alpha_{1} \mid 1
\rangle
$$
as a linear combination of $\mid 00000 \rangle $ and $ \mid 11111
\rangle$:
$$
\mid \varphi \rangle = \alpha_{0} \mid 00000 \rangle + \alpha_{1}
\mid 11111 \rangle.
$$
Alternatively, we can encode the qubit in state $ \mid \psi
\rangle$ as:

$$
\mid \varphi \rangle = \alpha_{0} \mid 0_{L} \rangle + \alpha_{1}
\mid 1_{L} \rangle,
$$
with $\mid 0_{L} \rangle$ and $ \mid 1_{L} \rangle$ expressed as a
superposition of codewords of a classical linear code. In this
case all codewords are superpositions of vectors in
$\mathcal{H}_{2^5}$, a Hilbert space of dimension $2^5$. Steane's
seven-qubit code \cite{Steane96a} and Shor's nine qubit code
\cite{Calderbank96} are based upon this scheme.

When we use the first encoding scheme, a random error can cause
departures from the subspace spanned by $\mid 00000 \rangle $ and
$\mid 11111 \rangle$. We should be able to correct small bit-flip
errors because the component which was $\mid 00000 \rangle$ is
likely to remain  in a sub-space $\mathcal{C}_{0} \subset
\mathcal{H}_{2^{5}}$ spanned by the six vectors:

$$
\mid 00000 \rangle, \mid 00001 \rangle, \mid 00010 \rangle, \mid
00100 \rangle, \mid 01000 \rangle, \mid 10000 \rangle
$$
while the component which was $\mid 11111 \rangle $ is likely to
remain in a sub-space $\mathcal{C}_{1} \subset
\mathcal{H}_{2^{5}}$ spanned by the six vectors,

$$\mid 11111 \rangle, \mid 11110 \rangle, \mid 11101 \rangle, \mid
11011 \rangle, \mid 10111 \rangle, \mid 01111 \rangle.
$$
The two subspaces are disjoint:

$$
\mathcal{C}_{0} \cap \mathcal{C}_{1} = \varnothing
$$
thus we are able to correct a bit-flip error of any single
physical qubit. This procedure is reminiscent of the basic idea of
classical error correction when we determine the Hamming sphere an
$n$-tuple belongs to, and then correct it as the codeword at the
center of the sphere.

A quantum code encodes one logical qubit into  $n$ physical qubits
in the Hilbert space $\mathcal{H}_{2^n}$ with basis vectors $\{
\mid 0 \rangle, \mid 1 \rangle, \ldots \mid i \rangle, \ldots \mid
2^{n} - 1 \rangle \}$. The two logical qubits are entangled states
in $\mathcal{H}_{2^n}$:

$$
\mid 0_{L} \rangle = \sum_{i} \alpha_{i} \mid i \rangle ~~~~~~~
\text{and}~~~~~~~
 \mid 1_{L} \rangle = \sum_{i} \beta_{i} \mid i \rangle
$$
A quantum error correcting code must map coherently the
two-dimensional Hilbert space spanned by $\mid 0_{L} \rangle$ and
$\mid 1_{L} \rangle$ into two-dimensional Hilbert spaces
corresponding to bit-flip, phase-flip, as well as bit-and-phase
flip of each of the $n$ qubits to ensure that the code is capable
of correcting the three types of error for each qubit. The {\it
quantum Hamming bound}:

$$
2 ( 3 n + 1) \le 2^{n}.
$$
established by Laflamme, Miquel, Paz, and Zurek in
\cite{Laflamme96} allows us to say that $n=5$ is the smallest
number of qubits required to encode the two superposition states
$\mid 0_{L} \rangle$ and $\mid 1_{L} \rangle$, and then be able to
recover them regardless of the qubit in error and the type of
errors.

The same paper \cite{Laflamme96} introduces a family of $5$-qubit
quantum error correcting codes. The code $\mathcal{Q}$ with the
logical codewords

$$
\begin{array} {cl}
\mid 0_L \rangle = \frac{1}{4} &
 (  \mid 00000 \rangle + \mid 10010 \rangle + \mid 01001 \rangle + \mid 10100 \rangle
  + \mid 01010 \rangle - \mid 11011 \rangle - \mid 00110 \rangle - \mid 11000 \rangle \\
 & - \mid 11101 \rangle - \mid 00011\rangle - \mid 11110 \rangle - \mid 01111 \rangle
   -\mid 10001\rangle-\mid 01100\rangle-\mid 10111\rangle+\mid
   00101\rangle ) \\\\
\mid 1_L\rangle=\frac{1}{4} &
 ( \mid 11111\rangle+\mid 01101\rangle+\mid 10110\rangle+\mid 01011\rangle+
  \mid 10101\rangle-\mid 00100\rangle-\mid 11001\rangle-\mid 00111 \rangle \\
 & -\mid 00010\rangle-\mid 11100\rangle-\mid 00001\rangle-\mid 10000\rangle-
 \mid 01110\rangle-\mid 10011\rangle-\mid 01000\rangle+\mid
 11010\rangle )
\end{array}
$$
is a member of this family. The code $\mathcal{Q}$ is a {\it
perfect quantum error correcting code} because a logical codeword
consists of  the smallest number $n$ of qubits to satisfy the
inequality from \cite{Laflamme96}. Recall that a {\it perfect
linear code} $[n, k, d]$ with $d=2e+1$ is one where Hamming
spheres of radius $e$ are all disjoint and exhaust the entire
space of $n$-tuples.

Different physical implementations reveal that the interactions of
the qubits with the environment are more complex and force us to
consider spatially- as well as, time-correlated errors. If the
qubits on an $n$ qubit register are confined to a $\tt {3D}$
structure, an error affecting one qubit will propagate to the
qubits in a volume centered around the qubit in error.
Spatially-correlated errors and means to deal with the spatial
noise are analyzed in \cite{Alicki02, Clemens04, Klesse05}. An
error affecting qubit $i$ of an $n$-qubit register may affect
other qubits of the register. An error affecting qubit $i$ at time
$t$ and corrected at time $t+\Delta$ may have further effect
either on qubit $i$ or on other qubits of the register. The error
model considered in this paper is based upon a recent study which
addresses the problem of reliable quantum computing using solid
state systems \cite{Novais06}.

There are two obvious approaches to deal with quantum correlated
errors :

\begin{itemize}
\item
 (i) design a code capable to correct these two or more errors, or
\item
(ii) use the classical information regarding past errors and
quantum error correcting codes capable to correct a single error.
\end{itemize}

When a quantum error correcting code uses a large number of qubits
and quantum gates it becomes increasingly more difficult to carry
out the encoding and syndrome measurements during a single quantum
error correction cycle. For example, if we encode a logical qubit
using Shor's nine qubit code we need $9$ physical qubits. If we
use a two-level convolutional code based upon Shor's code then we
need $81$ physical qubits and  $ 63 = 7 \times 9 $ ancilla qubits
to ensure that the circuit for syndrome calculation is
fault-tolerant. While constructing quantum codes capable of
correcting a larger number of errors is possible, we believe that
the price to pay, the increase in circuit complexity makes this
solution undesirable; this motivates our interest in the second
approach.

\section{Time-Correlated Quantum Errors}
\label{TimeCorrelatedErrors}

\begin{figure}[!ht]
\begin{center}
\includegraphics[width=12cm]{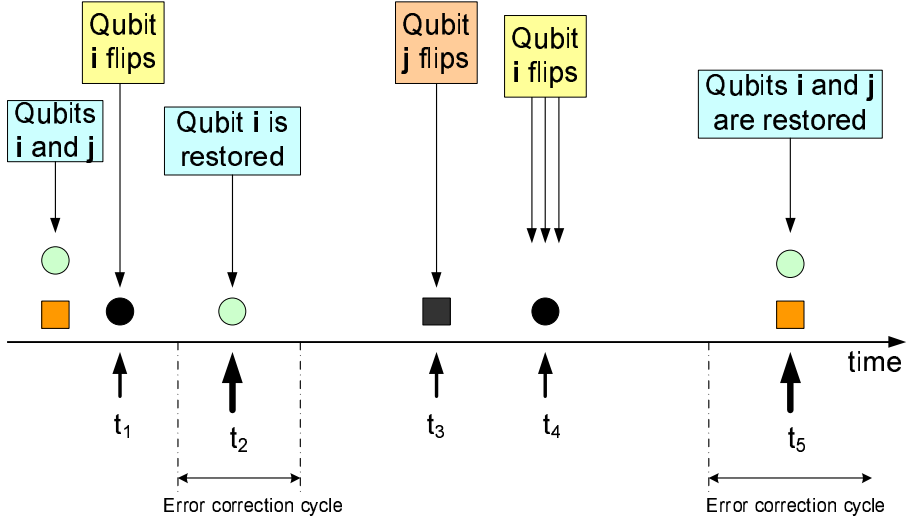}
\caption{Time-correlated quantum errors. Two consecutive error
correction cycles occurring at time $t_{2}$ and $t_{5}$  are
shown. At time  $t_{5}$ a code designed to correct a single error
will fail while a code capable to handle time-correlated errors
will correct the ``new'' error of qubit $j$ and the ``old''
correlated error of qubit $i$. } \label{TimeCorrErrs}
\end{center}
\end{figure}

Figure \ref{TimeCorrErrs} illustrates the evolution in time of two
qubits, $i$ and $j$ for two error correction cycles. The first
error correction cycle ends at time $t_{2}$ and the second at time
$t_{5}$. At time $t_{1}$ qubit $i$ is affected by decoherence and
flips; at time  $t_{2}$ it is flipped back to its original state
during the first error correction step; at time $t_{3}$ qubit $j$
is affected by decoherence and it is flipped; at time $t_{4}$ the
correlation effect discussed in this section affects qubit $i$ and
flips it to an error state. If an algorithm is capable to correct
one correlated error in addition to a ``new'' error, then during
the second error correction the errors affecting qubits $i$ and
$j$ are corrected at time $t_{5}$.

The quantum computer and the environment are entangled during the
quantum computation. When we measure  the state of the quantum
computer this entanglement is translated into a probability
$\epsilon$ that the measured state differs from the expected one.
This probability of error determines the actual number of errors a
quantum code is expected to correct.

If $\tau_{gate}$ is the time required for a single gate operation
and $\tau_{dch}$ is the decoherence time of a qubit, then
$n_{gates}$, the number of gates that can be traversed by a
register before it is affected by decoherence is given by:

$$
n_{gates} = { \tau_{dch} \over \tau_{gate}}.
$$

\begin{table}
\begin{center}
\caption{The time required for a single gate operation,
$\tau_{gate}$, the decoherence time of a qubit, $\tau_{dch}$, and
the number of gates that can be traversed before a register of
qubits is affected by decoherence, $n_{gates}$. }
\label{decoherenceTime}
\begin{tabular} {|r|lll|}
\hline
Qubit implementation & $\tau_{dch(sec)}$  & $\tau_{gate(sec)}$ & $n_{gates}$ \\
\hline
Nuclear spin          & $10^{4}$              & $10^{-3}$     & $10^{7} $   \\
Trapped Indium ion    & $10^{-1}$             & $10^{-14}$    & $10^{13}$   \\
Quantum dots/charge   & $10^{-9}$             & $10^{-12}$     & $10^{3}$    \\
Quantum dots/spin     & $10^{-6}$             & $10^{-9}$     & $10^{3}$    \\
Optical cavity        & $10^{-5}$             & $10^{-14}$    & $10^{9}$    \\
\hline
\end{tabular}
\end{center}
\end{table}

Quantum error correction is intimately related to the physical
processes which cause decoherence. Table \ref{decoherenceTime}
presents sample values of the time required for a single gate
operation $\tau_{gate}$, the decoherence time of a qubit,
$\tau_{dch},$ and $n_{gates}$, the number of gates that can be
traversed before a register of qubits is affected by decoherence,
for several qubit implementations
\cite{Hayashi03,Langer05,Petta05,Vandersypen01}. We notice a fair
range of values for the number of quantum gate operations that can
be performed before decoherence affects the state.

The information in Table \ref{decoherenceTime}, in particular the
decoherence time, can be used to determine the length of an error
correction cycle, the time elapsed between two consecutive error
correction steps. The number of quantum gate operations limits the
complexity of the quantum circuit required for quantum error
correction.

The Quantum Error Correction theory is based upon the assumption
that the quantum system has a constant error rate $\epsilon$. This
implies that once we correct an error at time $t_{c}$, the system
behavior at time $t > t_{c}$ is decoupled from events prior to
$t_{c}$.

The Markovian error model is not consistent with some physical
implementations of qubits; for example, in a recent paper Novais
and Baranger \cite{Novais06} discuss the decoherence in a
spin-boson model which is applicable, for instance, to quantum
dots. The authors assume that the qubits are perfect, thus
eventual errors are due to dephasing and consider a linear
coupling of the qubits to an ohmic bath. Using this model the
authors analyze the three-qubit Steane's code. They calculate the
probability of having errors in quantum error correction cycles
starting at times $t_{1}$ and $t_{2}$ and show that the
probability of errors consists of two terms; the first is the
uncorrelated probability and the second is the contribution due to
correlation between errors in different cycles ($\Delta$ is the
period of the error correcting cycle):

$$
P \approx ( { \epsilon \over 2} )^2 +  { { \lambda^{4} \Delta^4 }
\over { 8(t_{1}  - t_{2})^4 }}
$$
We see that correlations in the quantum system decay algebraically
in time, and the latest error will re-influence the system with a
much higher probability than others.

Only phase-flip errors are discussed in detail in \cite{Novais06}.
The approach introduced by the authors is very general and can be
extended to include other types of errors. Nevertheless, for other
physical models, one does expect major departures from the results
presented in \cite{Novais06}.

\section{Stabilizer Codes}
\label{StabilizerCodes}

The stabilizer formalism is a succinct manner of describing a
quantum error correcting code by a set of quantum operators
\cite{Gottesman97}. We first review several concepts and
properties of stabilizer codes.

The 1-qubit Pauli group $\mathcal{G}_{1}$
 consists of the Pauli operators, $\sigma_{I}$, $\sigma_{x}$,
 $\sigma_{y}$, and $\sigma_{z}$ together with the multiplicative
 factors $\pm 1$ and $\pm i$:

 $$
 \mathcal{G}_{1} \equiv \{ \pm \sigma_{I}, \pm i \sigma_{I}, \pm
 \sigma_{x}, \pm i \sigma_{x}, \pm \sigma_{y}, \pm i \sigma_{y},
 \pm \sigma_{z}, \pm i \sigma_{z} \}.
 $$
 The generators of $\mathcal{G}_{1} $ are:

 $$
 \langle \sigma_{x}, \sigma_{z}, i\sigma_{I}
 \rangle.
 $$
 The n-qubit Pauli group $\mathcal{G}_{n}$
 consists of the $4^{n}$ tensor products of $\sigma_{I}$,
 $\sigma_{x}$, $\sigma_{y}$, and $\sigma_{z}$ with an overall phase
 of $\pm 1$ or $\pm i$. Elements of the group can be used to describe the error
 operators applied to an $n$-qubit register. The {\it weight} of
 such an operator in $\mathcal{G}_{n}$ is equal to the number of
 tensor factors which are not equal to $\sigma_{I}$. The stabilizer
 $S$ of code $\mathcal{Q}$ is a subgroup of the
n-qubit Pauli group, $ S \subset \mathcal{G}_{n}$. The generators
of the subgroup $S$ are:

$$
M = \{ {\bf M}_{1}, {\bf M}_{2} \ldots {\bf M}_{q}\}.
$$
The eigenvectors of the generators $\{ {\bf M}_{1}, {\bf M}_{2}
\ldots {\bf M}_{q}\}$ have special properties: those corresponding
to eigenvalues of $+1$ are the codewords of $\mathcal{Q}$ and
those corresponding to eigenvalues of $-1$ are codewords affected
by errors. If a vector $\mid \psi_{i} \rangle \in \mathcal{H}_{n}$
satisfies,

$$
{\bf M}_{j} \mid \psi_{i} \rangle = (+1) \mid \psi_{i} \rangle
~~~~~~\forall {\bf M_j} \in M
$$
then $\mid \psi_{i} \rangle$ is a codeword, $\mid \psi_{i} \rangle
\in \mathcal{Q}$. This justifies the name given to the set $S$,
any operator in $S$ stabilizes a codeword, leaving the state of a
codeword unchanged. On the other hand if:

$$
{\bf M}_{j} \mid \varphi_{k} \rangle = (-1) \mid \varphi_{k}
\rangle.
$$
then $\mid \varphi_{k} \rangle = {\bf E}_i \mid \psi_{k} \rangle$,
the state $\mid \varphi_{k} \rangle$ is a codeword $\mid \psi_{k}
\rangle \in \mathcal{Q}$ affected by error ${\bf E}_i$. The error
operators affecting codewords in $\mathcal{Q}$, $E = \{ {\bf
E}_{1}, {\bf E}_{2} \ldots \} $, are also a subgroup of the
n-qubit Pauli group

$$
E \subset \mathcal{G}_{n}.
$$
Each error operator ${\bf E}_{i}$ is a tensor product of $n$ Pauli
matrices. Its weight is equal to the number of errors affecting a
quantum work, thus the number of Pauli matrices other than
$\sigma_{I}$.

The coding space:

$$
Q= \{ \mid \psi_{i} \rangle \in \mathcal{H}_{n} ~~\text{such
that}~~ M_j \mid \psi_{i} \rangle = (+1) \mid \psi_{i} \rangle,
~~~~~ \forall M_j \in S~\}
$$
is the space of all vectors $ \mid \psi_{i} \rangle$ fixed by $S$

\medskip

It is easy to prove that $S$ is a stabilizer of a non-trivial
Hilbert subspace $V_{2^n} \subset \mathcal{H}_{2^n}$ if and only
if:

\begin{enumerate}
\item
 $S= \{ {\bf S}_{1}, {\bf S}_{2}, \ldots \}$ is an Abelian group:

 $$
 {\bf S}_{i} {\bf S}_{j} = {\bf S}_{j} {\bf S}_{i}, ~~\forall {\bf S}_{i}, {\bf S}_{j} \in S
 ~~ i \ne j.
 $$

\item
 The identity matrix multiplied by $-1$ is not in $S$:

 $$
 - (\sigma_{I}^{\otimes n}) \notin S.
 $$
\end{enumerate}

If ${\bf E}$ is an error operator, and ${\bf E}$ anti-commutes
with some element ${\bf M } \in S$, then ${\bf E}$ can be
detected, since for any $ \mid \psi_{i} \rangle \in Q$:

$$
{\bf M E } \mid \psi_{i} \rangle  = {-\bf E M } \mid \psi_{i}
\rangle = - {\bf  E} \mid \psi_{i} \rangle
$$

\noindent {\definition The normalizer $N(S)$ consists of operators
${\bf E} \in \mathcal{G}_{n}$ such that ${\bf E} {\bf S}_{i} {\bf
E}^{\dagger} \in S, \forall {\bf S}_{i} \in S$. The distance $d$
of a stabilizer code is the minimum weight of an element in
$N(S)-S$ with $N(S)$ the normalizer of $S$. An $[n,k,d]$
stabilizer code is an $[n,k]$ code stabilized by $S$ and with
distance $d$; $n$ is the length of a codeword, and $k$ the number
of information symbols.}

\medskip

An $[n,k,d]$ stabilizer code with the minimum distance $d= 2e+1$
can correct at most $e$ arbitrary quantum errors or $2e$ errors
whose location is well known. In \cite{Dawson06} this property of
a stabilizer code is used for syndrome decoding for optical
cluster state quantum computation.

Given an $[n,k,d]$ stabilizer code the cardinalities of the
stabilizer $S$ and of its generator $M$ are:

$$
\mid S \mid = 2^{n-k},~~~~\mid M \mid = n-k
$$
The error syndrome corresponding to the stabilizer ${\bf M}_{j}$
is a function of the error operator, ${\bf E}$, defined as:

$$
 f_{{\bf M}_{j}} ({\bf E})
:\mathcal{G} \mapsto {\bf Z}_{2}~~~~~~~
 f_{{\bf M}_{j}}({\bf E} )=
 \left\{
 \begin{array}{ll}
 0 & \text{if}~~ [{\bf M}_{j}, {\bf E}] = 0 \\
 1 & \text{if} ~~ \{ {\bf M}_{j}, {\bf E} \} = 0,
 \end{array}
 \right.
$$
where $[{\bf M}_{j}, {\bf E}]$ is the commutator and $\{ {\bf
M}_{j}, {\bf E} \}$ the anti-commutator of operators ${\bf M}_{j}$
and ${\bf E}$. Let $f({\bf E} ) $ be the $(n-k)$-bit integer given
by the binary vector:

$$
f({\bf E} ) = (f_{{\bf M}_{1}}({\bf E}) f_{{\bf M}_{2}}({\bf E})
 \ldots f_{{\bf M}_{n-k}}({\bf E} )).
$$
This $(n-k)$-bit integer is called the {\it syndrome of error}
${\bf E}$.

\noindent {\proposition The error syndrome uniquely identifies the
qubit(s) in error if and only if the subsets of the stabilizer
group which anti-commute with the error operators are
distinct.}

\medskip

An error can be identified and corrected only if it can be
distinguished from any other error in the error set. Let $Q(S)$ be
the stabilizer code with stabilizer $S$. The \emph{Correctable Set
of Errors} for Q(S) includes all errors which can be detected by
$S$ and have distinct error syndromes.

\noindent {\corolary Given a quantum error correcting code $Q$
capable to correct $e_{u}$ errors, the syndrome does not allow us
to distinguish the case when more than $e_{u}$ qubits are in
error. If we can identify the $e_{c}$ correlated errors in
sysntem, the code is capable of correcting these $e_{u}+e_{c}$
errors. }

\medskip

\noindent {\it Proof: } Assume that $F_1$, $F_2$ cause at most
$e_{u}$ qubits to be in error, thus $F_1$, $F_2$ are included in
the \emph{Correctable Set of Errors} of $Q$. Assuming errors $F_1$
and $F_2$ are distinguishable, there must exist some operator $M
\in S$ which commutes with one of them, and anti-commutes with the
other:

\begin{center}
{$F_1^T F_2M=-MF_1^TF_2$}
\end{center}

If we know the exact correlated errors $E$ in the system, then:

$(E^TF_1)^T(E^TF_2)M=(F_1^TEE^TF_2)M=F_1^TF_2M=-MF_1^TF_2=-M(E^TF_1)^T(E^TF_2)$

\noindent which means that the stabilizer $M$ commutes with one of
the two errors $E^TF_1$, $E^TF_2$ and anti-commutes with the
other. So error $E^TF_1$ is distinguishable from error $E^TF_2$.
Therefore, if we know the exact prior errors $E$, we can identify
and correct any $E^TF_i$ errors with the weight of $F_i$ equal or
less than $e_{u}$.
\bigskip

For example, consider a $5$-qubit quantum error-correcting code
$\mathcal{Q}$ with $n=5$ and $k=1$ from \cite{Laflamme96}
discussed in  Section \ref{QuantumErrorCorrection}. The stabilizer
$S$ of this code is described by a group of 4 generators:

$$
\begin{array} {lll}
M= \{ {\bf M}_{1}, {\bf M}_{2}, {\bf M}_{3}, {\bf M}_{2} \}
  & \text{with the generators:} & \\
 &
 {\bf M}_{1} = \sigma_{x} \otimes \sigma_{z} \otimes \sigma_{z}
 \otimes \sigma_{x} \otimes \sigma_{I}, &
 {\bf M}_{2}=\sigma_{I} \otimes \sigma_{x} \otimes \sigma_{z} \otimes
 \sigma_{z} \otimes \sigma_{x}, \\
 &
 {\bf M}_{3} =\sigma_{x} \otimes \sigma_{I} \otimes \sigma_{x} \otimes
 \sigma_{z} \otimes \sigma_{z}, &
 {\bf M}_{4} =\sigma_{z} \otimes \sigma_{x} \otimes \sigma_{I} \otimes
 \sigma_{x} \otimes \sigma_{z}.
\end{array}
$$
It is easy to see that two codewords are eigenvectors of the
stabilizers with an eigenvalue of $(+1)$:

$$
 {\bf M}_{j} \mid 0_{L} \rangle = (+1) \mid 0_{L} \rangle
 ~~~~~\text{and}~~~~
 {\bf M}_{j} \mid 1_{L} \rangle = (+1) \mid 1_{L} \rangle,~~~1 \le
 j \le 4.
$$
Note also that $M$ is an Abelian subgroup, each generator commutes
with all the others. \medskip

\begin{table}
\caption{Single error operators for the 5-qubit code and the
generator(s) that anti-commute with each operator. Subscripts
indicate positions of errors, e.g. $X_3$ means a bit-flip error on
the third qubit, $X_3=\sigma_{I} \otimes \sigma_{I} \otimes
\sigma_{x} \otimes \sigma_{I} \otimes \sigma_{I}$.}
 \label{ErrorsAndGenerators}
$$
\begin{array} {|c|c|c|c|c|c|} \hline
 \text{Error operator} & \text{Generator(s)}& \text{Error operator} &
 \text{Generator(s)}& \text{Error operator} & \text{Generator(s)}\\ \hline

 {\bf X}_{1}  &
 {\bf M}_{4} &
 {\bf Z}_{1}  &
  {\bf M}_{1},~
  {\bf M}_{3} &
  {\bf Y}_{1}  &
  {\bf M}_{1},~
  {\bf M}_{3},~
  {\bf M}_{4}  \\

 {\bf X}_{2}  &
 {\bf M}_{1} &
 {\bf Z}_{2}  &
 {\bf M}_{2},~
  {\bf M}_{4} &
  {\bf Y}_{2}  &
{\bf M}_{1},~
 {\bf M}_{2},~
  {\bf M}_{4} \\

 {\bf X}_{3}  &
 {\bf M}_{1},~
 {\bf M}_{2} &
 {\bf Z}_{3}  &
 {\bf M}_{3}  &
 {\bf Y}_{3}  &
 {\bf M}_{1},~
 {\bf M}_{2}, ~
  {\bf M}_{3}\\

 {\bf X}_{4}  &
 {\bf M}_{2}, ~
 {\bf M}_{3}  &
 {\bf Z}_{4}  &
 {\bf M}_{1} ,~
 {\bf M}_{4} &
  {\bf Y}_{4}  &
 {\bf M}_{1} ,~
 {\bf M}_{2}, ~
 {\bf M}_{3}, ~
 {\bf M}_{4}  \\

 {\bf X}_{5}  &
 {\bf M}_{3}, ~
 {\bf M}_{4} &
 {\bf Z}_{5}  &
 {\bf M}_{2} &
 {\bf Y}_{5}  &
 {\bf M}_{2} ,~
 {\bf M}_{3}, ~
 {\bf M}_{4}\\
\hline

\end{array}
$$
\end{table}

Table  \ref{ErrorsAndGenerators} lists single error operators and
the generator(s) which anti-commute with each operator. For
example, ${\bf X}_{1}$ anti-commutes with ${\bf M}_{4}$, thus a
bit-flip on the first qubit can be detected; ${\bf Z}_{1}$
anti-commutes with ${\bf M}_{1}$ and ${\bf M}_{3}$, thus a
phase-flip of the first qubit can also be detected. Since each of
these 15 errors anti-commute with distinct subsets of $S$ we can
distinguish individual errors and then correct them. An example
shows that the code cannot detect two qubit errors;  indeed, the
two bit-flip error

$$
{\bf X}_{1} {\bf X}_{2} = \sigma_{x} \otimes \sigma_{x} \otimes
\sigma_{I} \otimes \sigma_{I} \otimes \sigma_{I}
$$
is indistinguishable from $ {\bf Z}_{4}=\sigma_{I} \otimes
\sigma_{I} \otimes \sigma_{I}
 \otimes \sigma_{z} \otimes \sigma_{I}$ because both ${\bf
X}_{1} {\bf X}_{2}$ and $ {\bf Z}_{4}$  anti-commute with the same
subset of stabilizers, $\{{\bf M}_{1},{\bf M}_{4}\}$, and give the
same error syndrome. Therefore, {\it the 5-qubit code can correct
any single qubit error, but cannot correct two qubit
errors}.\medskip

If we know the correlated error in the system, for example, ${\bf
X}_{3}$, then, from Table \ref{ErrorsAndGenerators}, it is easy to
see that the errors ${\bf X}_{3} {\bf E}_{i}$ have distinct error
syndromes for $1\leq i\leq 15$. Therefore we can identify these
errors and correct them.

\section{Time-Correlated Errors and Stabilizer Codes}
\label{StabilizerTimeCorrelatedErrors}

Classical, as well as quantum error correction schemes allow us to
construct codes with a well-defined error correction capability.
If a code is designed to correct $e$ errors, it will fail whenever
more than $e$ errors occur.

From previous section we know that if time-correlated errors are
present in the system, we can not always detect them through the
calculation of the syndrome. The same syndrome could signal the
presence of a single error, two, or more errors, as shown by our
example in Section \ref{StabilizerCodes}; we can remove this
ambiguity when there are two errors and one of them is a
time-correlated error.

In this section we extend the error correction capabilities of any
stabilizer code designed to correct a single error (bit-flip,
phase-flip error, or bit-and-phase flip) and allow the code to
correct an additional time-correlated error. The standard
assumptions for quantum error correction are:

\begin{itemize}

\item
Quantum gates are perfect and operate much faster than the
characteristic response of the environment;

\item The states of the computer can be prepared with no errors.

\end{itemize}

\noindent We also assume that:

\begin{itemize}

\item
There is no spatial-correlation of errors, a qubit in error does
not influence its neighbors;

\item
In each error correcting cycle, in addition to a new error $E_a$
that occurs with a constant probability $\varepsilon _a$, a
time-correlated error $E_b$  may occur with probability
$\varepsilon _b(t)$. As correlations decay in time, the qubit
affected by error during the previous error correction cycle, has
the highest probability to relapse.

\end{itemize}

Quantum error correction requires {\it non-demolition measurements
of the error syndrome} in order to preserve the state of the
physical qubits. In other words, a measurement of  the probe (the
ancilla qubits) should  not influence the free motion of the
signal system. The syndrome has to identify precisely the qubit(s)
in error and the type of error(s). Thus, the qubits of the
syndrome are either in the state with density $\mid 0 \rangle
\langle 0 \mid$ or in the state with density $\mid 1 \rangle
\langle 1 \mid$, which represent classical information.

A quantum non-demolition measurement allows us to construct the
error syndrome $\Sigma_{current}$. After examining the syndrome
$\Sigma_{current}$ an error correcting algorithm should be able to
decide whether:

\begin{enumerate}
\item
No error has occurred; no action should be taken;

\item
One ``new'' error, $E_a$, has occurred; then we apply the
corresponding Pauli transformation to the qubit in error;

\item
Two or more errors have occurred. There are two distinct
possibilities: (a) We have a ``new'' error as well as an ``old''
one, the time-correlated error; (b) There are two or more ``new''
errors. A quantum error correcting code capable of correcting a
single error will fail in both cases.

\end{enumerate}

It is rather hard to distinguish the last two possibilities. For
perfect codes, the syndrome $S_{ab}$ corresponding to two errors,
$E_a$ and $E_b$, is always identical to the syndrome $S_{c}$ for
some single error $E_c$. Thus,  for perfect quantum codes the
stabilizer formalism does not allow us to distinguish two errors
from a single one; for a non-perfect code it is sometimes possible
to distinguish the two syndromes and then using the knowledge
regarding the time-correlated error it may be possible to correct
both the ``old'' and the ``new'' error.

We now describe an algorithm based on the stabilizer formalism
capable to handle case 3(a) for perfect as well as non-perfect
quantum codes. We assume that the quantum code has a codeword
consisting of $n$ qubits and uses $k$ ancilla qubits for syndrome
measurement.

\medskip

\textbf{The outline of our algorithm:}

\begin{enumerate}

\item
At the end of an error correction cycle entangle the qubit
affected by the error with two additional ancilla qubits. Thus, if
the qubit relapses, the time-correlated error will propagate to
the two additional ancilla qubits.

\item
Carry out an {\it extended syndrome measurement} of the codeword
together with the two additional ancilla qubits. This syndrome
measurement should not alter the state of the codeword and keep
the entanglement between the codeword and the two additional
ancilla qubits intact.

\item
Disentangle the two additional ancilla qubits from the codeword and then measure the
two additional ancilla qubits.

\item
Carry out the error correction according to the outcomes of Steps
2 and 3.
\end{enumerate}

Next we use the Steane code to illustrate the details of the
algorithm. The generators of the Steane 7-qubit code are:

$$
\begin{array} {ll}
 {\bf M}_{1} = \sigma_{I}\otimes\sigma_{I}\otimes\sigma_{I}\otimes\sigma_{x} \otimes \sigma_{x} \otimes \sigma_{x}
 \otimes \sigma_{x}, &
 {\bf M}_{2}=\sigma_{I}\otimes\sigma_{x}\otimes\sigma_{x}\otimes\sigma_{I} \otimes \sigma_{I} \otimes \sigma_{x}
 \otimes \sigma_{x}, \\
 {\bf M}_{3} =\sigma_{x}\otimes\sigma_{I}\otimes\sigma_{x}\otimes\sigma_{I} \otimes \sigma_{x} \otimes \sigma_{I}
 \otimes \sigma_{x}, &
 {\bf M}_{4} =\sigma_{I}\otimes\sigma_{I}\otimes\sigma_{I}\otimes\sigma_{z} \otimes \sigma_{z} \otimes \sigma_{z}
 \otimes \sigma_{z}, \\
{\bf
M}_{5}=\sigma_{I}\otimes\sigma_{z}\otimes\sigma_{z}\otimes\sigma_{I}
\otimes \sigma_{I} \otimes \sigma_{z}
 \otimes \sigma_{z}, &
{\bf M}_{6}
=\sigma_{z}\otimes\sigma_{I}\otimes\sigma_{z}\otimes\sigma_{I}
\otimes \sigma_{z} \otimes \sigma_{I}
 \otimes \sigma_{z}.
\end{array}
$$

\begin{figure}[h]
\begin{center}
\includegraphics[width=14cm]{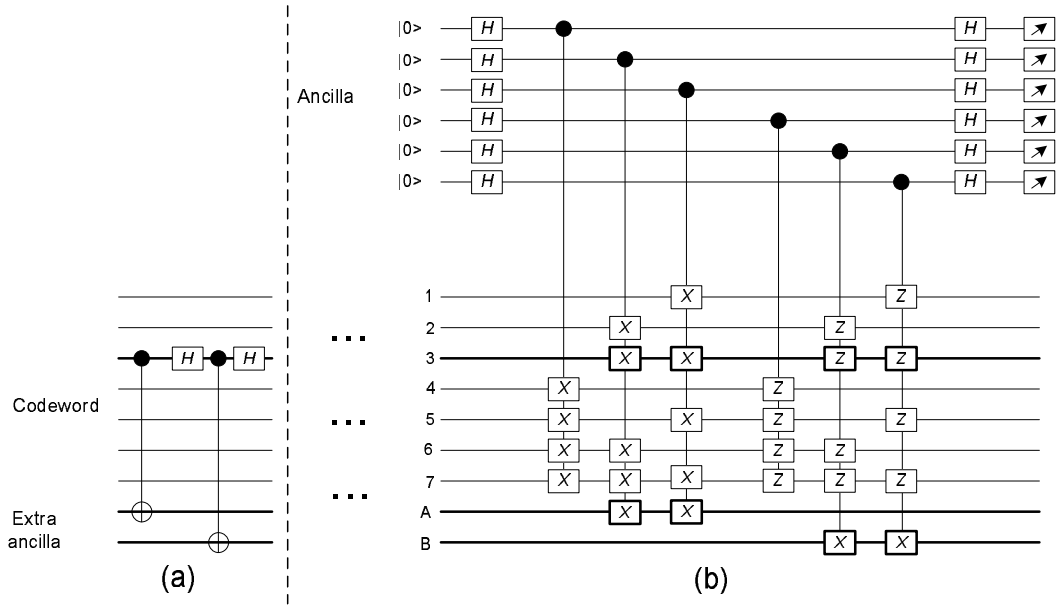}
\end{center}
\caption{(a) Duplicate the qubit affected  error during the last
cycle in both \textbf{X} and \textbf{Z} basis; (b) Extended
syndrome measurement on both the codeword and the additional
ancilla qubits.} \label{extended}
\end{figure}

\noindent {\bf 1. Entangle the two additional ancilla qubits with
the original code word.} The time-correlated error may reoccur in
different physical systems differently: a bit-flip could lead to a
bit-flip, a phase-flip, or a bit-phase-flip. Therefore, we need to
``backup'' the qubit corrected during the previous error
correction cycle in both \textbf{X} and \textbf{Z} basis. We
entangle the qubit affected by an error with two additional
ancilla qubits: a {\tt CNOT} gate will duplicate the qubit in
\textbf{Z} basis using the first ancilla qubit, the second ancilla
will duplicate the qubit in \textbf{X} basis using two Hadamard
gates and a {\tt CNOT} gate (called an {\tt HCNOT} gate), see
Figure \ref{extended}(a). In this example, qubit 3 is the one
affected by error in the last error correction cycle.

\medskip

\noindent {\bf 2. Extended syndrome measurement.} Measure the
extended syndrome  $S_{n+2}$ for an $(n+2)$ extended codeword
consisting of the original codeword and the two extra ancilla
qubit. Call $\Sigma$ the result of the measurement of the extended
syndrome $S_{n+2}$. When implementing this extended syndrome
measurement, we need consider two aspects: (i) the entanglement
between the additional ancilla and codewords should not be
disturbed, since we need to disentangle the extra ancilla from the
code and keep the codeword intact; and (ii) the new stabilizers
(or generators) should also stabilize the codeword and satisfy all
the stabilizers' requirements. In our approach, the additional
ancilla qubit \textbf{A} is a copy of the control qubit in
\textbf{Z} basis. To keep the entanglement, we copy all \textbf{X}
operations on the control qubit to qubit \textbf{A}. Similarly,
the additional ancilla qubit \textbf{B} is a copy of the control
qubit in \textbf{X} basis. We replicate all \textbf{Z} operations
performed on the control qubit to this qubit. Since it is in
\textbf{X} basis, the \textbf{Z} operations will change into
\textbf{X} operations. Consider the example of the Steane code and
the time-correlated error on qubit 3,  Figure \ref{extended}(b).
In this case we replicate all \textbf{X} operations performed on
qubit 3 to ancilla qubit \textbf{A}, and replicate all \textbf{Z}
operations to additional ancilla qubit \textbf{B} as \textbf{X}
operations.  These operations will keep the entanglement intact
during the syndrome measurement process. Also, it is easy to check
that the new stabilizers (i) commute with each other, and (ii)
stabilize the codewords.

\medskip

\begin{figure}[h]
\begin{center}
\includegraphics[width=10cm]{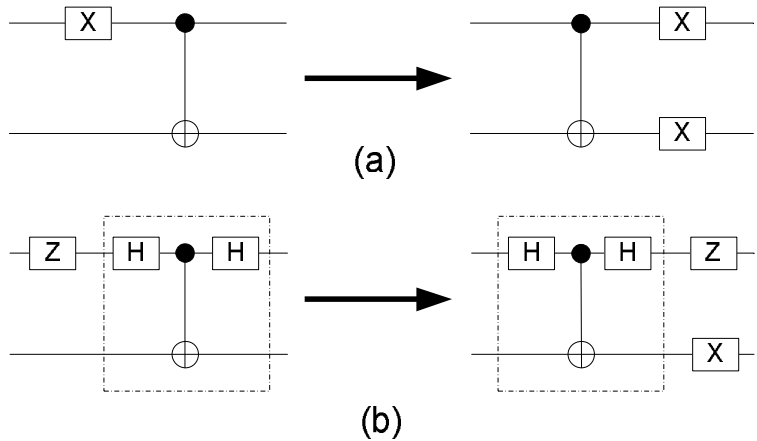}
\end{center}
\caption{(a) A {\tt CNOT} gate propagates a bit-flip of the
control qubit to the target qubit (b) An  {\tt HCNOT} gate
propagates a phase-flip on the control qubit to the target qubit
as a bit-flip.} \label{CNOT}
\end{figure}

\noindent {\bf 3. Disentangle and measure the two additional
ancilla qubits.} After the syndrome measurement, we use a circuit
similar to the one in Figure \ref{extended}(a) to disentangle the
two additional ancilla qubits from the codeword. The gates in this
circuit are in reverse order as in Figure \ref{extended}(a). Since
the entanglement is not affected during the extended syndrome
measurement, this process will disentangle the extra ancilla from
the codewords and keep the codeword in its original state. If
qubit 3 is affected by the time-correlated error, bit-flip,
phase-flip or both, the disentanglement circuit will propagate
these errors to the extra ancilla qubits: a {\tt CNOT} gate will
propagate a bit-flip of the control qubit to the target, and an
{\tt HCNOT} gate will propagate a phase-flip of the control qubit
to the target qubit as a bit-flip,  Figure \ref{CNOT}. Therefore,
measuring the additional ancilla qubits after the disentanglement
will give us the information regarding the type of error qubit 3
has relapsed to.

\medskip

\noindent {\bf 4. Correct the errors according to the output of
the extended syndrome measurement and of the measurement of the
two additional ancilla qubits.} Consider the following scenarios:

\begin{enumerate}
\item A ``new'' error affects a qubit $i$ of the codeword, while
qubit $j, j \ne i$ was the one in error during the previous cycle.
Then the measurement of the extended syndrome produces the same
outcome as the measurement of the original syndrome and the new
error is identified.

\item The time-correlated error of qubit $j$ occurs and there is
no ``new'' error. The error propagates to the two ancilla  qubits.
The measurement of the original syndrome allows us to identify the
single error.

\item There are two errors, the time-correlated error of qubit $j$
and a ``new'' error affecting qubits $i, i \ne j$ of the codeword.
Then the extended syndrome measurement gives the combination of
these two errors: $\Sigma = \Sigma_{j} \oplus \Sigma_{i}$. The
measurement of the two additional ancilla qubits reveals the type
of the time-correlated error: \emph{10} -- bit-flip, \emph{01} --
phase-flip, \emph{11} -- bit-and-phase-flip,  see Figure
\ref{CNOT}. Knowing that qubit $j$ is in error and the
corresponding error type we use the syndrome table to find the new
error on qubit $i$.

\item There is a single ``new'' error affecting one of the two
extra ancilla qubits. This error may propagate to the qubit of the
codeword the ancilla is entangled with. We treat this scenario as
(1). For example, in Figure \ref{extended}, a phase-flip of the
ancilla qubit \textbf{A} propagates to qubit 3. The measurement of
the original syndrome allows us to identify this error on qubit 3.

\item There are two errors, the time-correlated error on qubit $j$
of the codeword and an error on the additional ancilla entangled
with qubit $j$. As we mention in scenario (3), the error on extra
ancilla may back propagate to qubit $j$. The net result is that in
the codeword, only the qubit $j$ is affected by error: bit-flip,
phase-flip or bit-and-phase-flip. This error is indicated by the
syndrome and can be corrected as in scenario (1).

\end{enumerate}

The error correction process can be summarized as follows:

\begin{enumerate}
\item If the syndrome $\Sigma$ indicates an error on qubit $j$,
the one affected by error during the last cycle (qubit 3 in our
example), correct the error;

\item
Else, determine  the outcome of a measurement of the two
additional ancilla qubits:

\begin{enumerate}
\item
If the outcome is \emph{00}, and $\Sigma$ corresponds to a single
error syndrome, correct this error;

\item If the outcome is either \emph{01}, or \emph{10}, or
\emph{11}, qubit $j$ in error during the previous cycle has
relapsed. Then the syndrome is $\Sigma_{i}=\Sigma \otimes
\Sigma_{j}$. Knowing that qubit $j$ is in error and the
corresponding error type (\emph{10} -- bit-flip, \emph{01} --
phase-flip, \emph{11} -- bit-and-phase-flip) we use the syndrome
table to find the new error affecting qubit $i$.

\item Otherwise, there are two or more new errors in the system
and they cannot be corrected.

\end{enumerate}
\end{enumerate}

\noindent {\example Consider again the Steane sever-qubit code.}
Table \ref{syndromes} shows the syndromes for all single errors
for the Steane 7-qubit quantum code. Assume that during the last
error correction cycle qubit $j=3$ was affected by an error. We
use a {\tt CNOT} gate to entangle qubit 3 with the additional
ancilla qubit \textbf{A} in \textbf{Z} basis, and an {\tt HCNOT}
gate to entangle it with qubit \textbf{B} in \textbf{X} basis,
Figure \ref{extended}.

\begin{table}[h!]
 \caption{The syndromes for each of the three types of errors of each
qubit of a codeword for the Steane 7-qubit code: $X_{1-7}$
bit-flip, $Z_{1-7}$ phase-flip, and  $Y_{1-7}$ bit-and-phase flip.
}
\begin{center}
\begin{tabular} {c|c||c|c||c|c|}
\hline

Error & Syndrome & Error & Syndrome & Error & Syndrome\\
\hline

\hline $X_{1}$ &   000001 &

 $Z_{1}$  &  001000 &
 $Y_{1}$  &  001001\\

 \hline $X_{2}$ &  000010 &

  $Z_{2}$ &   010000&
  $Y_{2}$ &   010010\\

\hline $X_{3}$ &  000011 &

 $Z_{3}$ &  011000&
 $Y_{3}$ &  011011\\

\hline $X_{4}$ &   000100 &

 $Z_{4}$ &   100000&
  $Y_{4}$ &   100100\\

\hline  $X_{5}$ &   000101 &

 $Z_{5}$ &   101000&
 $Y_{5}$ &   101101\\

 \hline  $X_{6}$ &   000110 &

 $Z_{6}$ &   110000&
 $Y_{6}$ &   110110\\

 \hline  $X_{7}$ &   000111 &

 $Z_{7}$ &   111000&
 $Y_{7}$ &   111111\\

\hline
\end{tabular}
\end{center}
 \label{syndromes}
\end{table}

\begin{enumerate}
\item
$\Sigma = 011000$, indicates a phase-flip on qubit 3.
We correct this error. In this case the measurement of the
additional ancilla qubit could be non-zero, but this will not
affect the codeword.

\item
$\Sigma = 110000$ and the outcome of the measurement of the two
additional ancilla qubits is $00$. In this case there is only one
``new'' error in the system, a phase-flip on qubit $i=6$ as
indicated by $\Sigma$. We correct this error.

\item
$\Sigma = 110000$ and the outcome of the measurement of the
additional ancilla qubits is $01$. In this case
$\Sigma_{j}=\Sigma_{phase-flip}=\Sigma_{Z3}=011000$. Therefore,
the new error has syndrome $\Sigma_{i}=110000 \otimes 011000 =
101000$; this indicates  a phase-flip of qubit $i=5$. Correct the
phase-flips on qubit 3 and qubit 5.
\end{enumerate}

The circuits described in our paper have been tested using the
Stabilizer Circuit Simulator \cite{Anders06}; sample codes are
provided in the Appendix. One can test different error scenarios
using a code similar to the one in the Appendix.

\section{Summary}

Errors affecting the physical implementation of  quantum circuits
are not independent, often they are time-correlated. Based on the
properties of time-correlated noise, we present an algorithm that
allows the correction of one time-correlated error in addition to
a new error. The algorithm can be applied to any stabilizer code
when the two logical qubits $\mid 0_L \rangle$ and $\mid 1_L
\rangle$ are entangled states of $2^{n}$ basis states in
$\mathcal{H}_{2^n}$.

The algorithms can be applied to  perfect as well as non-perfect
quantum codes. The algorithm requires two additional ancilla
qubits entangled with the qubit affected by an error during the
previous error correction cycle. The alternative is to use a
quantum error correcting code capable of correcting two errors in
one error correction cycles; this alternative requires a
considerably more complex quantum circuit for error correction.

\section{Acknowledgments}

The research reported in this paper was partially supported by
National Science Foundation grant CCF 0523603. The authors express
their gratitude to Lov Grover, Daniel Gottesman and Eduardo
Mucciolo for their insightful comments.

\newpage
\oddsidemargin 0.0in

\begin{tabular}{p{3.25in}p{3.25in}}
\multicolumn{2}{p{6.5in}}{\begin{center}\bf APPENDIX \end{center}}\\
\scriptsize
\begin{quote}
\begin{verbatim}
# Sample code for testing the algorithm
# Requires graphsim from
# http://homepage.uibk.ac.at/~c705213/work/graphsim.html

import graphsim
import random random.seed()
gr=graphsim.GraphRegister(22,random.randint(0,1E6))

# Prepare Steane |0> codeword on qubit 0-6
gr.hadamard (4)
gr.hadamard (5)
gr.hadamard (6)
gr.cnot (6, 3)
gr.cnot (6, 1)
gr.cnot (6, 0)
gr.cnot (5, 3)
gr.cnot (5, 2)
gr.cnot (5, 0)
gr.cnot (4, 3)
gr.cnot (4, 2)
gr.cnot (4, 1)

# Entangle qubit 3 with the extra ancilla
gr.cnot(2,7)
gr.hadamard(2)
gr.cnot(2,8)
gr.hadamard(2)

# Errors happen on qubit ?
gr.bitflip(2)
gr.phaseflip(3)

##### Begin current error correction cycle #####

# "Extended" Syndrome extraction
for i in range(9,15):
      gr.hadamard(i)

# Z syndrome
gr.cnot(9,3)
gr.cnot(9,4)
gr.cnot(9,5)
gr.cnot(9,6)

gr.cnot(10,1)
gr.cnot(10,2)
gr.cnot(10,5)
gr.cnot(10,6)
gr.cnot(10,7)

gr.cnot(11,0)
gr.cnot(11,2)
gr.cnot(11,4)
gr.cnot(11,6)
gr.cnot(11,7)

# X syndrome
gr.cphase(12,3)
gr.cphase(12,4)
gr.cphase(12,5)
gr.cphase(12,6)

gr.cphase(13,1)
gr.cphase(13,2)
gr.cphase(13,5)
gr.cphase(13,6)
gr.cnot(13,8)

gr.cphase(14,0)
gr.cphase(14,2)
gr.cphase(14,4)
gr.cphase(14,6)
gr.cnot(14,8)

\end{verbatim}
\end{quote} &
\scriptsize
\begin{quote}
\begin{verbatim}
# "Disentangle" the extra ancilla
gr.hadamard(2)
gr.cnot(2,8)
gr.hadamard(2)
gr.cnot(2,7)

# Measure and output the extra ancilla
print gr.measure(7)
print gr.measure(8)

#gr.print_adj_list ()
#gr.print_stabilizer ()
print

# Measure and output syndrome
for i in range(9,15):
  gr.hadamard(i)
  print gr.measure(i)
print

#############   error correction  ###############

# Error correction according to the syndrome and extra ancilla
gr.bitflip(2)
gr.phaseflip(3)

##### Normal syndrome extraction to double-check  #####

for i in range(9+6,15+6):
  gr.hadamard(i)

# Z syndrome
gr.cnot(9+6,3)
gr.cnot(9+6,4)
gr.cnot(9+6,5)
gr.cnot(9+6,6)

gr.cnot(10+6,1)
gr.cnot(10+6,2)
gr.cnot(10+6,5)
gr.cnot(10+6,6)

gr.cnot(11+6,0)
gr.cnot(11+6,2)
gr.cnot(11+6,4)
gr.cnot(11+6,6)

# X syndrome
gr.cphase(12+6,3)
gr.cphase(12+6,4)
gr.cphase(12+6,5)
gr.cphase(12+6,6)

gr.cphase(13+6,1)
gr.cphase(13+6,2)
gr.cphase(13+6,5)
gr.cphase(13+6,6)

gr.cphase(14+6,0)
gr.cphase(14+6,2)
gr.cphase(14+6,4)
gr.cphase(14+6,6)

# Measure and output the syndrome
for i in range(9+6,15+6):
  gr.hadamard(i)
  print gr.measure(i)
print

\end{verbatim}
\end{quote}
\\

\end{tabular}

\end{document}